\documentclass[preprint]{aastex}

\RequirePackage{color}

\begin{document}

\title{Alfv\'en node-free vibrations of white dwarf in the model
of solid star with toroidal magnetic field}
\shorttitle{Alfv\'en vibrations of a solid star with toroidal magnetic field}
\shortauthors{Molodtsova, Bastrukov, Chen, Chang}

\author{Irina Molodtsova\altaffilmark{1}, Sergey Bastrukov\altaffilmark{1,2}, Kuan-Ting Chen\altaffilmark{2},
Hsiang-Kuang Chang\altaffilmark{2,3}}
\affil{National Tsing Hua University, Hsinchu 30013, Taiwan}
\affil{Joint Institute for Nuclear Research, 141980 Dubna, Russia}

\altaffiltext{1}{Joint Institute for Nuclear Research, 141980 Dubna, Russia}
\altaffiltext{2}{Institute of Astronomy, National Tsing Hua University, Hsinchu 30013, Taiwan}
\altaffiltext{3}{Department of Physics, National Tsing Hua University, Hsinchu 30013, Taiwan}

\begin{abstract}
 In the context of the white dwarf asteroseismology,
 we investigate vibrational properties of a non-convective solid star with an axisymmetric purely toroidal intrinsic magnetic field of two different shapes. Focus is laid on regime of node-free global Lorentz-force-driven vibrations about symmetry axis at which material displacements have one and the same form as those for nodeless spheroidal and torsional vibrations restored by Hooke's force of elastic shear stresses. Particular attention is given to the even-parity poloidal Alfv\'en modes whose frequency spectra are computed in analytic form showing how the purely toroidal magnetic fields completely buried beneath the star surface can manifest itself in seismic vibrations of non-magnetic white dwarfs. The obtained spectral formulae are discussed in juxtaposition with those
 for Alfv\'en modes in the solid star model with the poloidal, homogeneous internal and dipolar external, magnetic field whose inferences are relevant to Alfv\'en vibrations in magnetic white dwarfs.
\end{abstract}



\section{Introduction}
 An understanding peculiarities of Alfv\'en oscillations in solid star models is of particular interest for the asteroseismology of final stage (FS) compact objects of stellar evolution, white dwarfs and neutron stars.
 The super dense and highly conducting matter of these stars, capable of accumulating magnetic fields of extremely powerful intensity (e.g., Chanmugam 1992), possesses mechanical elasticity generic to solids. This suggests
 that proper account of seismic vibrations of FS stars can be attained working from equations of elastodynamics and magneto-solid-mechanics, contrary to the liquid stars of main sequence (MS) whose asteroseismology is adequately modeled by equations of hydrodynamics and magneto-fluid-mechanics. Another important feature distinguishing liquid MS stars from solid FS stars is that the magnetic fields of the MS stars are generated in the dynamo processes whose energy supply comes from central thermonuclear reactive zone (Parker 1979), whereas in the
 FS stars there are no nuclear energy sources to support persistent convection. It is commonly believed today
 that magnetic fields of non-convective FS stars are fossil (e.g. Ferrario and Wickramasinghe 2004). In these latter
 non-convective stars, like white dwarfs, the fossil magnetic fields are regarded as frozen in the immobile
 extremely dense and perfectly conducting matter.

 Most studies devoted to interpretation of quasi-periodic oscillations (QPOs) in the electromagnetic emission of FS stars as being produced by Lorentz-force-driven Alfv\'en oscillations have been perform within the framework of the fiducial homogeneous model of a solid star with an axisymmetric poloidal internal magnetic field of constant intensity
 \begin{eqnarray}
  \label{e1.1}
  && {\bf B}({\rm PMF})=[B_r=\cos\theta\,\,B_\theta=-\sin\theta,\,B_\phi=0],\\
  \label{e1.2}
  && W_m({\rm PMF})=\frac{1}{8\pi}\int B^2\,d{\cal V}=\frac{1}{6}B^2R^3
  \end{eqnarray}
  where $W_m$ is the internal magnetic energy of the star. In particular, in our recent papers (Bastrukov et al 2009a, 2009b) this model has been invoked to investigate a poorly studied regime of global node-free Alfv\'en oscillations in quaking neutron stars. Based on the well-known argument that magnetic field frozen-in a perfectly conducting matter imparts to stellar material a supplementary portion of elasticity (e.g. Chandrasekhar 1961), and taking into account formal similarity between equations of classical elastodynamics and magneto-solid-mechanics, it has been conjectured that material displacements in a neutron star undergoing node-free oscillations restored by Lorentz force of magnetic field stresses should have one and the same form as those for a solid star executing nodeless spheroidal and torsional shear oscillations restored by Hooke's elastic force of solid-mechanical stresses whose frequency spectra has early been computed in (Bastrukov et al 2007, 2008). Following this line of argument
  it has been shown that this suggestion and the energy variational method of magneto-solid-mechanical theory allow one to get exact spectral formulae for the frequencies of the node-free poloidal and toroidal $a$-modes in neutron stars.

  In the present paper we discuss application of magneto-solid-mechanical theory of Alfv\'en oscillations developed in (Bastrukov et al 2009b) to the white dwarf asteroseismology. The material of this wide class of pulsating solid stars is identified with a metal-like solid-state plasma composed of ions suspended in the degenerate Fermi-gas of relativistic electrons. Therefore if one assume that all pulsating white dwarfs, both magnetic and non-magnetic,
  come into existence with frozen-in magnetic fields, then there seems to be no theoretical reason why the
  phenomenon of Alfv\'en vibrations should not occur in these stars. To the best of our knowledge,
  the possibility that magnetic DA and DB white dwarfs endowed with poloidal magnetic fields (having dipolar-like external configuration) are capable of maintaining Alfv\'en oscillations has profoundly been discussed some time ago by Lou (1995) who was the first to present quite convincing arguments that global Alfv\'en oscillations, entrapped
  in the entire volume of white dwarf, can be self-excited via  $\kappa$-mechanism of ionization (of hydrogen in DA and helium in DB stars), that is, by one and the same mechanism which
  is responsible of excitation in white dwarfs of $g$-modes (Winget 1998, Hansen, Kawaler and  Trimble 2004, Fontain and Brassard 2008, Winget and Kepler 2008).

  Adhering to this line of argument, in this article we investigate peculiarities of Lorentz-force-driven global spheroidal and torsional nodeless vibrations of a solid star with frozen-in purely toroidal and completely intrinsic magnetic field ${\bf B}({\rm TMF})=[B_r=0,B_\theta=0,B_\phi\neq 0]$ and discuss expected observational consequences of node-free poloidal and toroidal Alfv\'en oscillations in non-magnetic white dwarfs endowed with toroidal
  magnetic field completely buried beneath its surface.
  First, is the model with the azimuthal component of magnetic field having the form
   \begin{eqnarray}
  \label{e1.3}
  && B_\phi({\rm TMF}1)=B \frac{R-r}{R}\sin\theta
  \end{eqnarray}
   which is pictured, with the aid of computer code MAPLE, in Fig.1.

   \begin{figure}[h]
\centering\
\includegraphics[width=5.0cm]{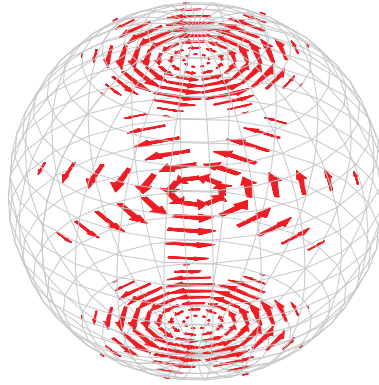}
\caption{
The lines of toroidal magnetic field $B_\phi({\rm TMF1})=B[(R-r)/R]\sin\theta$ entrapped in the star volume.}
\end{figure}

  And second is the model with pictured in Fig.2. azimuthal component of magnetic field of the form
   \begin{eqnarray}
  \label{e1.4}
  && B_\phi({\rm TMF2})=B \frac{[R^2-r^2\sin^2\theta]^{1/2}}{R}.
  \end{eqnarray}
   It may be worth noting that the store of magnetic energy in the stars with the exactly above forms of toroidal magnetic fields is less than that for a star model with poloidal homogeneous internal and dipolar external
   magnetic field\footnote{It is easily seen, however, that replacement of constant field $B$ by $\tilde B=kB$ in the expressions for $B_\phi({\rm TMF}1)$ and $B_\phi({\rm TMF}2)$, with constant value of $k$ equal to $\sqrt{15}$ for the former model and $\sqrt{5/3}$ for the latter one, leads to energy of magnetic field precisely coinciding  with that for a star with uniform internal poloidal magnetic field.}
     \begin{eqnarray}
  \label{e1.5}
  && W_m({\rm TMF}1)=\frac{1}{90}B^2R^3=\frac{1}{15}W_m({\rm PMF}),\\
   && W_m({\rm TMF2})=\frac{1}{10}B^2R^3=\frac{3}{5}W_m({\rm PMF}).
  \end{eqnarray}
The practical usefulness of such models is that their inferences can be utilized in the asteroseismology of wide class of pulsating non-magnetic white dwarfs if we are to understand the impact of toroidal magnetic field, entrapped in the volume of the star, on seismic vibrations restored by Lorentz force of magnetic field stresses and to identify characteristic spectral features of these vibrations in the detected QPOs of X-ray and optical emission from these stars.

\begin{figure}[h]
\centering\
\includegraphics[width=5.0cm]{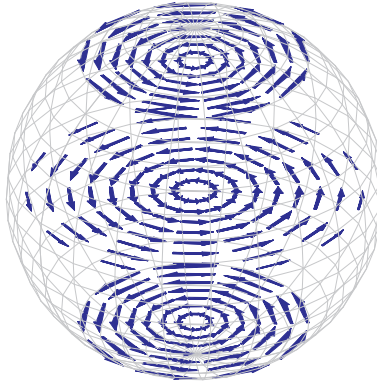}
\caption{
The same as Fig.1 but for  $B_\phi({\rm TMF2})=(B/R)[R^2-r^2\sin^2\theta]^{1/2}$.}
\end{figure}

   It is worth noting that node-free Alfv\'en oscillations in polytropic self-gravitating gaseous masses (governed by equations of magneto-fluid-mechanics lying at the base of asteroseimology of liquid MS stars) with purely toroidal magnetic field have been the subject of long-lasting investigations in the past (Chandrasekhar 1956, Anand \& Kushwaha R.S. 1962, Roxburgh 1962, Roxburgh and Durney 1967, Anand 1969, Sood and Trehan 1972; Miketinac 1973; Goossens 1976, Goossens and Veugelen 1978, Goossens and Biront 1980). In most of these focus was on
   a liquid star model with purely toroidal magnetic field of the form, $B_\phi=(B/R)r\sin\theta$. The net outcome of these studies is that the presence of such a field can significantly modify the frequencies of spheroidal
   non-radial nodeless oscillations such as in Kelvin fundamental mode (Bastrukov et al 2009c).
   In the meantime, the regime of nodeless Alfv\'en oscillations in the solid FS stars with purely toroidal magnetic field  remains less studied. In Sec.2, we show that proper account of this case can done on the basis of the
   energy method of magneto-solid-mechanical theory that has been extensively discussed in our recent work
   (Bastrukov et al 2009b). In Sections 3 and 4, based on this method the detailed calculation of the frequency spectra of node-free poloidal and toroidal $a$-modes in a solid stars with intrinsic purely azimuthal magnetic fields of above form is presented and discussed in the context of current development of the white dwarf asteroseismology. The newly obtained results are briefly summarized in Sec.5.

 \section{Brief outline of governing equations}

 The governing equations of solid-magnetics (formulated as a solid-mechanical counterpart of equations
 of hydromagnetics) describing perfectly conducting elastic continuous medium of density $\rho$ pervaded by magnetic field $B_i$ are
 \begin{eqnarray}
  \label{e2.1}
  &&\rho{\ddot u}_i=\nabla_k\,\delta M_{ik},\\
 \label{e1.2}
  &&\delta M_{ik}=\frac{1}{4\pi}[B_i\delta B_k+B_k\delta B_i -B_j\delta B_j \delta_{ik}],\\
  \label{e1.3}
  &&\delta B_i=\nabla_k\,[u_i\,B_k-u_k\,B_i],\\
  && \frac{\partial }{\partial t} \int \frac{\rho {\dot u}^2}{2}\,
 d{\cal V}=-\int \delta M_{ik}\,{\dot u}_{ik}\,d{\cal V}.
 \end{eqnarray}
  As in (Bastrukov et al 2009a, 2009b, 2009c) in what follows we consider vibrations in which material displacements ${\bf u}$ are described by harmonic solenoidal fields obeying vector Laplace equation
  \begin{eqnarray}
  \label{e2.4}
  \nabla^2 {\bf u}=0,\quad \nabla\cdot {\bf u}=0
  \end{eqnarray}
  which is regarded as characteristic equation of the regime of node-free shear vibrations\footnote{In the context of the solid-mechanical treatment of non-compressional, $\nabla_k u_k=0$, seismic stellar vibrations, governed by
  equation $\rho{\ddot {\bf u}}=\nabla_k\sigma_{ik}$ with shear stresses obeying Hooke's law $\sigma_{ik}=2\mu\,u_{ik}$ with $\mu$ being the shear modulus of star matter, the Laplace equation can be thought of as the long-wavelength limit of Helmholtz equation, $\nabla^2 {\bf u}+ k^2{\bf u}=0$. This latter equation emerges from above solid-dynamical equation after substitution $u_i=u_i^0\exp i({\bf kr}-\omega t)$ and describes standing-wave regime of elastic shear vibrations. Understandably that
  in the limit of long waves $\lambda=(2\pi/k)\to \infty$, i.e. when wave vector $k\to 0$, the Helmholtz equation is reduced
  to the Laplace equation (Bastrukov et al 2007).}.

  Bearing in mind the above mentioned similarity between oscillatory behavior of magneto-active solid-state plasma and
  elastically deformable solid, the field of instantaneous, i.e. time-independent, material displacements,  $a_i({\bf r})$, which is related to $u_i({\bf r},t)$ as
 \begin{eqnarray}
 \label{e2.5}
 && u_i({\bf r},t) =a_i({\bf r}){\alpha}(t)
 \end{eqnarray}
 has been been taken in one and the same form of poloidal ${\bf a}_p$ and toroidal ${\bf a}_t$ vector fields
 as in the case of nodeless spheroidal and torsional shear oscillations restored by Hooke's force of elastic shear stresses (Bastrukov et al 2007, 2008)
 \begin{eqnarray}
  \label{e2.6}
  && {\bf a}_p({\bf r})=A_p\nabla\times\nabla\times [{\bf r}\chi({\bf r})],\\
  \label{e2.7}
  && {\bf a}_t({\bf r})=A_t\nabla\times [{\bf r}\chi({\bf r})],\\
  \label{e2.8}
  && \chi({\bf r})=r^\ell\,P_\ell(\zeta),\quad \zeta=\cos\theta
  \end{eqnarray}
  where $P_\ell(\cos\theta)$ is the Legendre polynomial of multipole degree $\ell$ and to accentuate the absence nodes in radial dependence of these fields in the interval $[0\leq r\leq R]$ (from the star center, $r=0$, to its surface $r=R$), the term node-free vibrations is used.
  These fields are basic ingredients of parameters of inertia ${\cal M}$ and stiffness of magneto-mechanical rigidity ${\cal K}_m$ of the Hamiltonian of Alfv\'en oscillations
  \begin{eqnarray}
 \label{e2.9}
 {\cal H}=\frac{{\cal M}{\dot \alpha^2(t)}}{2}+\frac{{\cal K}_m{\alpha}^2(t)}{2}.
 \end{eqnarray}
 The explicit expressions form of these parameters read
 \begin{eqnarray}
 \label{e2.10}
 && {\cal M}=\int \rho\, a_i\,a_i d{\cal V},\quad
 {\cal K}_m=\int \tau_{ik}({\bf r})\,a_{ik}({\bf r})\,d{\cal V}
 \end{eqnarray}
 where
 \begin{eqnarray}
 \label{e2.11}
 && \tau_{ik}({\bf r})=\frac{1}{4\pi}[B_i({\bf r})\,b_k({\bf r})+B_k\,({\bf r})b_i({\bf r})],\\
 \label{e2.12}
 && b_i({\bf r})=(B_k ({\bf r})\nabla_k)a_i({\bf r})-(a_k({\bf r})\nabla_k) B_i({\bf r}),\\
 \label{e2.13}
 && a_{ik}({\bf r})=\frac{1}{2}[\nabla_i\,a_k({\bf r})+\nabla_k\,a_i({\bf r})].
  \end{eqnarray}
  Thus, knowing the density profile $\rho$ and space distribution
  of equilibrium magnetic field $B_i$ in the vibrating star, the above outlined energy method provides a unified way of computing frequencies $\omega^2={\cal K}_m/{\cal M}$ of both poloidal and toroidal modes of Alfv\'en node-free oscillations.

  Following this line of argument, in work (Bastrukov et al 2009a), the results of which are of particular interest for our discussion here, it was found that the frequency spectral of poloidal and toroidal $a$-modes are given by
 \begin{eqnarray}
 \label{e.14}
 && \omega(_0a^p_\ell)=\omega_A\left[\ell(\ell-1)\frac{2\ell+1}{2\ell-1}\right]^{1/2},\\
 \label{e2.15}
 && \omega(_0a^t_\ell)=\omega_A\left[(\ell^2-1)\frac{2\ell+3}{2\ell-1}\right]^{1/2},\\
 \label{e2.16}
 && \omega_A=\frac{v_A}{R},\quad v_A=\frac{B}{\sqrt{4\pi\rho}},\\
 \label{e2.17}
 &&\omega_A=B\sqrt{\frac{R}{3M}},\quad M=\frac{4\pi}{3}\rho R^3.
 \end{eqnarray}
  Hereafter $v_A$ and $\omega_A$ stand for the Alfv\'en velocity and frequency, respectively,  $M$ is the mass and $R$ is the radius of a star.
  The asymptotic behavior of above spectra, at very high overtones, $\ell>>1$, is given by
  $\omega(_0a_\ell)/\omega_A\to \ell$ and similar asymptotic behavior exhibits frequency spectrum of
  toroidal Alfv\'en mode in solid star with Ferraro's form of non-homogeneous poloidal magnetic field (Bastrukov et 2009b). These spectral formulae provide us information about spacing between overtones of nodeless oscillations which is crucial to identification of $a$-modes in spectra of observed QPOs in the emission of magnetic white dwarfs. In the reminder of this work we focus on axisymmetric Alfv\'en oscillations in a solid star model with frozen-in toroidal magnetic fields presented which can be relevant to the asteroseismology of non-magnetic white dwarfs.

 \section{Torsional Alfv\'en oscillations in a solid star with toroidal magnetic field}

  Computation of integral parameters ${\cal M}$ and ${\cal K}_m$ for both TMF1 and TMF2 solid star models undergoing
  torsional vibrations about axis of symmetry with toroidal field of differentially rotational displacements
 \begin{eqnarray}
  \label{e2.1}
  && {\bf a}_t({\bf r})=A_t\nabla\times [{\bf r}\,r^\ell\,P_\ell(\zeta)],\\
  \nonumber
 &&a_r=0,\quad a_\theta=0,\\
 \nonumber
 &&  a_{\phi}=A_t\,r^\ell(1-\zeta^2)^{1/2}P'_\ell(\zeta),\quad
 P'_\ell(\zeta)=\frac{dP_\ell(\zeta)}{d\zeta}
  \end{eqnarray}
  yields
  \begin{eqnarray}
 \label{e2.2}
 && {\cal M}=4\pi\rho  A_t^2 R^{2\ell+3}\frac{\ell(\ell+1)}{(2\ell+1)(2\ell+3)},\quad {\cal K}_m=0.
 \end{eqnarray}
  The impossibility of sustaining differentially rotational vibrations about axis of symmetry is one
  of the basic findings of this study that hardly could be foreseen without straightforward computations.
  From Hamiltonian of the eigenfrequency problem in question
  it follows that quake-induced perturbation of such displacements should lead to
  differential rotation with kinetic energy $E={\cal M}{\dot \alpha}^2/2$, not vibrations, as is the case of
  solid star with poloidal magnetic field.

  \section{Spheroidal Alfv\'en oscillations in a solid star with toroidal magnetic field}

  For the former, it worth noting that in the star undergoing axisymmetric spheroidal vibrations under consideration
  the instantaneous material displacements are described by nodeless poloidal field of instantaneous material displacements which is irrotational
  \begin{eqnarray}
  \label{e3.1}
  && {\bf a}^p=A_p\nabla\times\nabla\times [{\bf r}\chi({\bf r})]={\cal A}_p\nabla r^\ell\,P_\ell(\cos\theta),\\
  \label{e3.2}
  && {\cal A}_p=\frac{A_p}{\ell+1},\quad \nabla\times  {\bf a}^p=0,\\
 \nonumber
 && a_r={\cal A}_p\,r^{\ell-1}\ell P_\ell(\zeta),\\
 \nonumber
 && a_\theta=-{\cal A}_p\,r^{\ell-1}(1-\zeta^2)^{1/2}P'_\ell(\zeta),\quad a_\phi=0.
 \end{eqnarray}
 The distortions of lines of toroidal magnetic field in the solid star undergoing axisymmetric spheroidal nodeless
 vibrations are pictured in Fig.3.

\begin{figure}[h]
\centering\
\includegraphics[width=8.0cm]{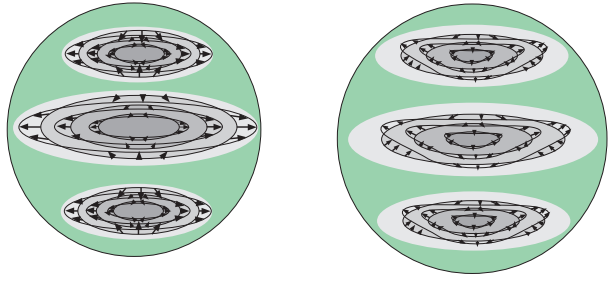}
\caption{\small
 Distortions of lines of toroidal magnetic field in the solid star undergoing spheroidal axisymmetric
 nodeless  vibrations in quadrupole (left) and octupole (right) overtones of poloidal Alfv\'en mode.}
\end{figure}

 The mass parameter of nodeless spheroidal oscillations ${\cal M}$ can conveniently be represented in the form
\begin{eqnarray}
 \nonumber
 M&=&\int \rho(r)
 [a_r^2+a_\theta^2] d{\cal V}\\
\label{e3.3}
 &=&4\pi\rho\,{\cal A}_p^2
 R^{2\ell+1}\,m_\ell,\quad m_\ell=
 \frac{\ell}{2\ell+1}
\end{eqnarray}
Understandably that computed mass parameter has one and the same value for both B(TMF1)
and B(TMF2) models and because the distinction between frequency spectra can only be due to
the difference in ${\cal K}_m$, we consider computation of this parameter in some details.

\subsection{B(TMF1)-model}

 Using computational formulae of above energy method reported in our previous work (Bastrukov et al 2009b),
 one finds that components of instantaneous distortions of magnetic field lines, $b_i({\bf r})=(B_k \nabla_k)a_i-(a_k\nabla_k) B_i$, are given by
 \begin{eqnarray}
 \nonumber
 && b_r=0,\quad b_\theta=0,\\
 \label{e3.4}
 && b_\phi=
{\cal A}_p\,B\,r^{\ell-2}\,\ell\,(1-\zeta^2)^{1/2}\,P_\ell(\zeta).
 \end{eqnarray}
 The tensor of perturbation-induced magnetic field stresses, $\tau_{ik}=(1/4\pi)[B_i\,b_k+B_k\,b_i]$, has only one non-zero component
 \begin{eqnarray}
  \nonumber
 \tau_{\phi\phi}&=&\frac{1}{2\pi}[B_\phi\,b_\phi]\\
 \label{e3.5}
 &=&\frac{{\cal A}_p\,B^2}{2\pi}\,
\frac{r^{n-2}(R-r)}{R}\,
\ell(1-\zeta^2)\,P_\ell(\zeta)
 \end{eqnarray}
so that the integrand of ${\cal K}_m$ reads
\begin{eqnarray}
 \label{e3.6}
 \tau_{\phi\phi} a_{\phi\phi}&=&\frac{{\cal A}_p^2\,B^2}{2\pi}
 \frac{r^{2\ell-4}(R-r)}{R}\,\ell\,\\ \nonumber
 &\times& (1-\zeta^2)\,P_\ell(\zeta)\left[\ell P_\ell(\zeta)-\zeta\,P'_\ell(\zeta)\right].
 \end{eqnarray}
Integration over spherical polar coordinates $r$ and $\theta$ yields
\begin{eqnarray}
  \nonumber
 && {\cal K}_m=\int [\tau_{\phi\phi}\,a_{\phi\phi}]d{\cal V}=\frac{{\cal A}_p^2\,B^2\,R^{2\ell-1}}{2\ell(2\ell-1)}
 \ell\left[\ell\, I_1- I_2\right],
 \end{eqnarray}
 where
 \begin{eqnarray}
   \nonumber
 &&I_1=\int\limits_{-1}^{1} (1-\zeta^2) P_\ell(\zeta) P_\ell(\zeta) d\zeta
     =\frac{4\,(\ell^2+\ell-1)}{(4\ell^2-1)(2\ell+3)},
 \end{eqnarray}
 and
 \begin{eqnarray}
  \nonumber
 &&I_2=\int\limits_{-1}^{1}\zeta(1-\zeta^2)P_\ell(\zeta)\,P'_\ell(\zeta) d\zeta
     =\frac{2\,\ell\,(\ell+1)}{(4\ell^2-1)(2\ell+3)}
\end{eqnarray}
Finally, we obtain
\begin{eqnarray}
 \label{e3.7}
 {\cal K}_m={\cal A}_p^2\,B^2\,R^{2\ell-1}\, k_\ell,\quad k_\ell=\frac{\ell\,(\ell-1)}{(2\ell+1)(2\ell-1)^2}
\end{eqnarray}
The frequency spectrum is given by
\begin{eqnarray}
 \label{e3.8}
 && \omega^2({\rm TMF1})=\omega^2_A\,\left[\frac{(\ell-1)}{(2\ell-1)^2}\right].
 \end{eqnarray}

  \begin{figure}[ht]
\centering\
\includegraphics[width=8.0cm]{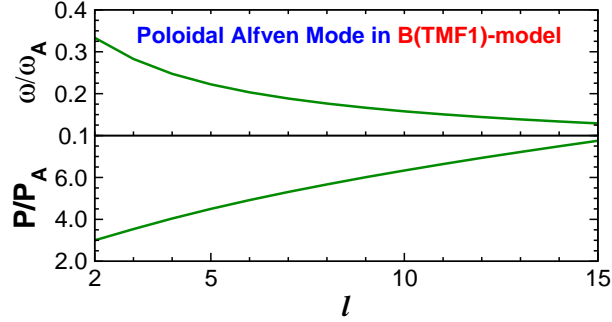}
\caption{\small
 Fractional frequencies and periods as functions of multipole degree of spheroidal Alfv\'en oscillations
 in the TMF1-model of a solid star with azimuthal component of toroidal magnetic fields, $B_\phi({\rm TMF1})=B[(R-r)/R]\sin\theta$.}
\end{figure}

 In Fig. 4 we plot fractional frequencies and
 periods as functions of multipole degree of axisymmetric spheroidal Alfv\'en nodeless oscillations
 of solid star in its own toroidal magnetic field of the TMF1-model.

 \subsection{B(TMF2)-model}

 Following the same line of argument, in this case of TMF2-model for
 instantaneous distortions of magnetic field lines $b_i$ we obtain
 \begin{eqnarray}
 \label{e4.1}
 && b_r=0,\quad b_\theta=0,\\
 \nonumber
 && b_\phi=-
{\cal A}_p\,B\,
\frac{R\,r^{\ell-2}}{\sqrt{R^2-r^2\,(1-\zeta^2)}}\,\zeta\,P'_\ell(\zeta).
 \end{eqnarray}
 As in previous case the tensor of $\tau_{ik}$ has too only one non-zero component
 \begin{eqnarray}
  \label{e4.2}
  \tau_{\phi\phi}=\frac{1}{2\pi}[B_\phi\,b_\phi]=-\frac{{\cal A}_p\,B^2}{2\pi}\,
r^{n-2}\,
\zeta\,P'_\ell(\zeta)
 \end{eqnarray}
and explicit form of integrand of ${\cal K}_m$ reads
\begin{eqnarray}
 \label{e4.3}
 \tau_{\phi\phi} a_{\phi\phi}&=&\frac{{\cal A}_p^2\,B^2\,r^{2\ell-4}}{2\pi}
 \\ \nonumber
  &\times&\zeta\,P'_\ell(\zeta)
\,\left[\zeta\,P'_\ell(\zeta)-\ell P_\ell(\zeta)\right].
 \end{eqnarray}
The integral for stiffness taken over $r$ and $\theta$ coordinates is given by
\begin{eqnarray}
 \nonumber
 && {\cal K}_m=\int [\tau_{\phi\phi}\,a_{\phi\phi}]d{\cal V}=\frac{{\cal A}_p^2\,B^2R^{2\ell-1}}{(2\ell-1)}
\,\left[ I_3- \ell\,I_4\right],\\
 \nonumber
 &&I_3=\int\limits_{-1}^{1} \zeta^2 \,P'^2_\ell(\zeta) d\zeta
     =\frac{\ell\,(\ell+1)\,(2\ell-1)}{(2\ell+1)},\\
 \nonumber
 &&I_4=\int\limits_{-1}^{1}\zeta\,P_\ell(\zeta)\,P'_\ell(\zeta) d\zeta
     =\frac{2\,\ell}{(2\ell+1)}
\end{eqnarray}
and its resultant expression reads
\begin{eqnarray}
 \label{e4.4}
{\cal K}_m={\cal A}_p^2\,B^2\,R^{2\ell-1}\, k_\ell,\quad
 k_\ell=\frac{\ell\,(\ell-1)}{(2\ell-1)}.
\end{eqnarray}

\begin{figure}[ht]
\centering\
\includegraphics[width=8.0cm]{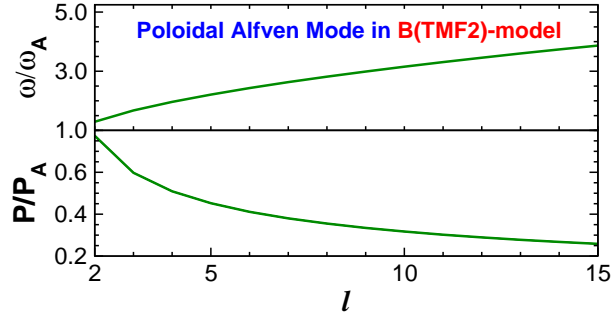}
\caption{\small
 The same as fig.4 but in the solid star with toroidal magnetic fields $B({\rm TMF2})=(B/R)[R^2-r^2\sin^2\theta]^{1/2}$.}
\end{figure}

The frequency spectrum of poloidal Alfv\'en mode in the star with magnetic field of TMF2-model
is given by
\begin{eqnarray}
 \label{e4.5}
 && \omega^2({\rm TMF2})=\omega_A^2(\ell-1)\,\frac{(2\ell+1)}{(2\ell-1)}.
\end{eqnarray}
This spectrum is visualized in Fig. 5.
 It follows from frequency spectra that the lowest overtone is of quadrupole multipole degree, $\ell=2$, as should be the case
 in view of shear character of vibrations at which density remains unaltered. The excitation of
 dipole, $\ell=1$, displacements results in center-of-mass translation motion of the star, as follows from Hamiltonian\footnote{It worth noting that computation of magneto-mechanical rigidity
 with above mentioned form of the azimuthal component of toroidal field $B_\phi=(B/R)r\sin\theta$ leads to ${\cal K}_m=0$. Thus, the fact that the toloidal magnetic field vanishes outside (completely
 entrapped in the star body) is crucial for the star capability of maintaining non-radial spheroidal vibrations with nodeless field of displacements in question.}.
For typical of white dwarfs parameters (the average density of degenerate matter $\rho\sim 10^5-10^6$ g/cm$^3$, the radius $R\sim 10^9$ cm and the intensity of intrinsic magnetic field $B\sim 10^8$ G), the basic period of poloidal Alfv\'en mode $P_A=\nu_A^{-1}$, where $\nu_A=\omega_A/(2\pi)$ falls in the range
$3\cdot 10^2\leq P_A \leq 5\cdot 10^3$ sec, that is, in the realm of  periods of quasi-periodic oscillations
of optical and X-ray emission detected form pulsating non-magnetic white dwarf stars.

\section{Concluding remarks}

 The fact that extremely dense metal-like matter of white dwarf stars, composed of ions suspended in the Fermi-gas of relativistic electrons, is capable of accommodating magnetic field of very high intensity, suggests that these solid stars could be one of the best objects for studying Alfv\'en oscillations (Lou 1995). Therefore the possibility should not be overlooked that, by not considering such oscillations, we can miss discovering some of important features of the white dwarf asteroseismology.

  \begin{figure}[ht]
 \centering\
 \includegraphics[width=7.30cm]{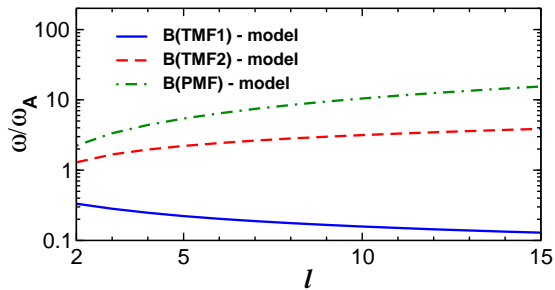}
 \caption{\small
 Fractional frequencies as a function of multipole degree of Lorentz-force-driven node-free
 poloidal Alfv\'en node-free oscillations, in a solid star models with
 toroidal magnetic fields, ${\bf B}({\rm TMF1})=[B_r=0,B_\theta=0,B_\phi=B[(R-r)/R]\sin\theta]$, ${\bf B}({\rm TMF2})=[B_r=0,B_\theta=0,B_\phi=(B/R)[R^2-r^2\sin^2\theta]^{1/2}]$ in juxtaposition with those for
 a solid star model with homogeneous poloidal magnetic field
 ${\bf B}({\rm PMF})=[B_r=B\cos\theta ,B_\theta=-B\sin\theta,B_\phi=0]$.}
\end{figure}

 Motivated by this line of argument and continuing our investigations of node-free regime of seismic vibrations of degenerate FS solid stars (Bastrukov et al 2007-2009), in this paper focus was laid on search for vibrational
 peculiarities of node-free Alfv\'en modes in the solid star model with purely toroidal magnetic field entrapped in the star volume. One of the main purposes was to get a feeling about differences between frequency spectra of global Alfv\'en node-free vibrations in the magnetic white dwarf (model of solid star endowed with axisymmetric poloidal, homogeneous internal and dipolar, external magnetic field) and non-magnetic white dwarf (solid star model with purely toroidal and completely internal magnetic field). What is newly disclosed here is that non-magnetic white dwarf with purely toroidal magnetic field can sustain only non-rotational spheroidal oscillations. And we found that asymptotic behavior of frequencies of poloidal Alfv\'en modes in a solid star with frozen-in purely azimuthal field is substantially different from
 that in the con-convective star with homogeneous poloidal magnetic field.
Specifically, for B(TMF1)-model, appropriate for a non-magnetic white dwarf, we have found
\begin{eqnarray}
 \label{e5.1}
 \frac{\omega({\rm TMF1})}{\omega_A}=\left[\frac{(\ell-1)}{(2\ell-1)^2}\right]^{1/2}\to \frac{1}{\sqrt{4\ell}},\quad
  \ell>>1.
\end{eqnarray}
For a solid star with azimuthal magnetic field of the B(TMF2)-type we found
\begin{eqnarray}
 \label{e5.3}
\frac{\omega({\rm TMF2})}{\omega_A}=\left[(\ell-1)\,\frac{(2\ell+1)}{(2\ell-1)}\right]^{1/2}\to \sqrt{\ell},\quad
\ell>>1.
\end{eqnarray}
An absolutely different asymptotic is expected for node-free poloidal $a$-mode in a solid star with homogeneous
uniform magnetic field, the PMF-model appropriate for magnetic white dwarf
\begin{eqnarray}
 \label{e5.2}
 \frac{\omega({\rm PMF})}{\omega_A}=\left[\ell(\ell-1)\frac{2\ell+1}{2\ell-1}\right]^{1/2}\to \ell,\quad \ell>>1.
\end{eqnarray}
 This comparison is visualized in Fig.6. Numerically, the computed periods of $a$-modes
 are in the realm of the QPO periods observed in the optical and X-ray emission of pulsating white dwarfs.
 All the above suggests that considered $a$-modes of white dwarf vibrations
 can manifest itself  in one line with $g$-modes and this feature of the white dwarf asteroseismology worths, therefore, more detailed studying.

\acknowledgements We are indebted to referee for valuable suggestions.
 This work is a part of investigations on variability of high-energy emission from compact sources
 supported by NSC of Taiwan, grant numbers  NSC-96-2628-M-007-012-MY3 and NSC-97-2811-M-007-003.

\end{document}